\documentclass[12pt]{article}
\usepackage{amssymb,amsmath,epsfig}

\begin{document}

\title{\bf Analysis of Generalized Ghost Version of Pilgrim Dark Energy}
\author{M. Sharif$^1$ \thanks{msharif.math@pu.edu.pk} and Abdul Jawad$^{1,2}$
\thanks{jawadab181@yahoo.com}\\
$^1$ Department of Mathematics, University of the Punjab,\\
Quaid-e-Azam Campus, Lahore-54590, Pakistan.\\
$^2$ Department of Mathematics, Lahore Leads University,\\
Lahore, Pakistan.}

\date{}

\maketitle
\begin{abstract}
The proposal of pilgrim dark energy is based on the speculation that
phantom-like dark energy possesses enough resistive force to
preclude the black hole formation in the later universe. We explore
this phenomenon by assuming the generalized ghost version of pilgrim
dark energy. We find that most of the values of the interacting
($\xi^2$) as well as pilgrim dark energy ($u$) parameters push the
equation of state parameter towards phantom region. The squared
speed of sound shows that this model remains stable in most of the
cases of $\xi^2$ and $u$. We also develop
$\omega_\Lambda-\omega'_\Lambda$ plane and observe that this model
corresponds to thawing as well as freezing regions. Finally, it is
shown that the non-interacting and interacting generalized ghost
versions of pilgrim dark energy correspond to $\Lambda$CDM limit on
the statefinder plane.
\end{abstract}

\section{Introduction}

The observational analysis through different schemes indicates that
the universe undergoes an accelerated expansion. The first clue
about this expansionary phenomenon was given by a variety of
astronomers through Supernova type Ia (Riess et al. 1998; Perlmutter
et al. 1999) and subsequent attempts also favor this phenomenon
(Caldwell and Doran 2004; Koivisto and Mota 2006; Hoekstra and Jain
2008). This phenomenon is assumed to be taken place under the
influence of an exotic type force termed as dark energy (DE) whose
nature is still unknown. To peruse this problem, a versatile study
has been done in various ways which has led many dynamical DE models
(Armendariz-Picon et al. 1999; Caldwell 2002; Bagla et al. 2003; Hsu
2004; Li 2004; Feng et al. 2005; Zhang et al. 2006; Cai 2007) and
modified theories of gravity (Brans and Dicke 1961; Linder 2010;
Dutta and Saridakis 2010; Sharif and Rani 2013a; 2013b).

The dynamical DE models have been constructed in two frameworks:
quantum gravity and general relativity. The holographic dark energy
(HDE) model (Li 2004) has been developed in the context of quantum
gravity with the help of holographic principle (Susskind 1995). The
basic idea behind this model is that the bound on the vacuum energy
$(\Lambda)$ of a system with size $L$ should not cross the limit of
the BH mass having the same size due to the formation of BH in
quantum field theory (Cohen et al. 1999). Wei (2012) reconsidered
this idea and suggested that if we are able to prevent the BH
formation in the later universe then the energy bound proposed by
Cohen et al. (1999) could be violated. For this purpose, the strong
repulsive force is required which may help in the avoidance of
matter collapse and hence the BH formation.

The question arises which type of force can play this role? In this
regard, phantom-like DE can give useful contribution as compared to
its other versions (vacuum and quintessence-like). It is well-known
that due to phantom-like DE, everything will be crashed before our
universe ends in the big-rip and ultimately BH formation would be
avoided in this way. The effects of phantom-like DE have been
explored in different ways. Babichev et al. (2004) suggested that
when phantom-like DE accreted onto a BH, it leads to the loss of BH
mass with the passage of time. This phenomenon has been tested
through different phantom-like family of chaplygin gas models which
lead to similar results (Martin-Moruno 2008; Jamil et al. 2008;
Babichev et al. 2008; Jamil 2009; Jamil and Qadir 2011; Sharif and
Abbas 2011; Bhadra and Debnath 2012). In the wormhole physics,
phantom-like DE plays an effective role in precluding the formation
of event horizon (Lobo 2005a, 2005b; Sushkov 2005, Sharif and Jawad
2014).

Harada et al. (2006) argued that self-similar solution does not
exist for a universe filled with quintessence or scalar field or
stiff fluid. The same results have been found by using different
approaches (Akhoury et al. 2009). However, in these works, only the
strong energy condition is violated which is inferred as
quintessence-like DE and does not possess enough resistive force to
preclude the BH formation. It was also pointed out by Li and Wang
(2007) that BHs can exist in FRW universe in the presence of DE
which only satisfies weak energy condition. Recently, Wei (2012) has
suggested that phantom-like DE is more effective since it violates
all the energy conditions. He constructed the following model
\begin{equation}\label{1a}
\rho_{\Lambda}=3\delta^2m^{4-u}_{p}L^{-u},
\end{equation}
where $\delta$ and $u$ appear as dimensionless constants, $L$ is
known as infrared cutoff and $m_p$ is the reduced Planck constant.
This model is called pilgrim DE (PDE). He considered PDE with Hubble
horizon as an IR cutoff and pointed out (through theoretical and
observational ways) different possibilities for the avoidance of BH
in the later universe.

We have extended this work for non-interacting and interacting PDE
with different IR cutoffs in the flat as well as non-flat universe
models (Sharif and Jawad 2013a, 2013b). In these works, we have
found that EoS parameter for non-interacting and interacting cases
lies in the phantom region for $u>0$ as well as $u<0$ which favor
PDE phenomenon. We have also developed
$\omega_\Lambda-\omega'_\Lambda$ and $r-s$ planes for these models.
Here, we explore the non-interacting and interacting generalized
ghost PDE models in the flat universe. We analyze the behavior of
EoS parameter, squared speed of sound (for stability),
$\omega_\Lambda-\omega'_\Lambda$ and $r-s$ planes. The paper is
organized as follows. Section \textbf{2} is devoted to the basic
equations, EoS parameter and the stability analysis. In sections
\textbf{3} and \textbf{4}, we present the analysis of
$\omega_\Lambda-\omega'_\Lambda$ and $r-s$ planes, respectively. We
summarize our results in the last section.

\section{Equation of State Parameter}

In this section, we present basic scenario of interacting
generalized ghost DE version of PDE with cold dark matter (CDM) in
flat universe. We extract equation of state parameter and analyze
its behavior through PDE parameter. The first equation of motion
corresponding to flat universe leads to
\begin{equation}\label{1}
H^2=\frac{1}{3m^2_{pl}}(\rho_m+\rho_\Lambda),
\end{equation}
where $\rho_m$ and $\rho_\Lambda$ indicate CDM and DE densities. In
terms of fractional energy density, this equation becomes
\begin{equation}\label{2}
\Omega_m+\Omega_\Lambda=1, \quad
\Omega_{m}=\frac{\rho_m}{3m^2_{pl}H^2},\quad
\Omega_{\Lambda}=\frac{\rho_\Lambda}{3m^2_{pl}H^2}.
\end{equation}
The proposal of Veneziano ghost DE ($\rho_\Lambda=\alpha H$, where
$\alpha$ is a constant with dimension $[energy]^3$) lies in the
category of dynamical DE models which plays an important role in the
accelerated expansion of the universe (Urban and Zhitnitsky 2009a,
2009b, 2010a, 2010b, 2011). The motivation of this model comes from
Veneziano ghost of choromodynamics (QCD) which is useful to solve
$U(1)$ problem in QCD. The key feature of this model is that
Veneziano ghost (being unphysical in quantum field theory
formulation in the Minkowski spacetime) provides non-trivial
physical effects in FRW universe (Rosenzweig et al. 1980; Nath and
Arnowitt 1981). The QCD ghost has a little contribution in the
vacuum energy density proportional to $\Lambda^{3}_{QCD}H$ (where
$\Lambda_{QCD}\sim100MeV$ is the smallest QCD scale and $H$
represents the Hubble parameter), but this contribution is very
crucial in the evolutionary behavior of the universe. This model
helps in alleviating the fine tuning as well as cosmic coincidence
problem (Urban and Zhitnitsky 2009a, 2009b, 2010a, 2010b, 2011;
Forbes and Zhitnitsky 2008). Several theoretical aspects have been
investigated for this model (Ebrahimi and Sheykhi 2011; Sheykhi and
Sadegh 2012; Sheykhi and Bagheri 2011; Rozas-Fernandez 2012; Karami
and Fahimi 2013) and tested thorough different observational schemes
(Cai et al. 2011).

It is noted that vacuum energy from Veneziano ghost field in QCD is
of the form $H+O(H^2)$ (Zhitnitsky 2012), but in the ordinary ghost
DE model, only the leading term (i.e., $H$) has been considered. Cai
et al. (2012) suggested that the contribution of subleading term
(i.e., $H^2$) in the ordinary ghost DE can be helpful in describing
the early evolution of the universe. He proposed the so called
generalized ghost DE density
\begin{equation*}
\rho_{\Lambda}=\alpha H+\beta H^2,
\end{equation*}
where $\beta$ is another constant with dimension $[energy]^2$.
Theoretically, different cosmological parameters such as EoS
parameter, deceleration, $\omega_\Lambda-\omega'_\Lambda$ and
statefinders etc have been developed for this model (Ebrahimi et al.
2012; Malekjani 2013; Karami 2013). The stability of this model has
also been investigated (Ebrahimi and Sheykhi 2013). Here, we use
this model in order to discuss the PDE phenomenon. In terms of PDE,
the generalized ghost DE density takes the form
\begin{equation}\label{3}
\rho_{\Lambda}=(\alpha H+\beta H^2)^u,
\end{equation}
known as generalized ghost PDE.

We take interaction between generalized ghost version of PDE with
CDM which follows the equations of continuity as
\begin{eqnarray}\label{4}
\dot{\rho}_{m}+3H\rho_{m}=\Delta,\quad
\dot{\rho}_{\Lambda}+3H(\rho_{\Lambda}+p_{\Lambda})=-\Delta,
\end{eqnarray}
where $\Delta$ possesses dynamical nature and appears as interaction
term between CDM and generalized ghost DE version of PDE. In
general, three forms are commonly used given as follows
\begin{eqnarray*}\label{4}
\Delta_1&=&3\xi^2H\rho_{\Lambda},\\
\Delta_2&=&3\xi^2H(\rho_m+\rho_{\Lambda}),\\
\Delta_3&=&3\xi^2H\rho_{m}.
\end{eqnarray*}
Here $\xi^2$ is the coupling constant whose sign is important in the
present cosmological evolution of the universe. The positive $\xi^2$
represents the decay of DE into DM while negative $\xi^2$ is
responsible for decomposition of DM into DE. It is argued (Pavon and
Wang 2009) that the sign of $\xi^2$ should be taken as positive
according to thermodynamical view point as second law of
thermodynamics favors the decay of DE into DM.

Moreover, it was found that observations also support the phenemonon
of decaying of CDM into DE (Pereira and Jesus 2009; Guo, et al.
2007; Costa, et al. 2009). Cai and Su (2010) fitted the interaction
term $\Delta$ with observations without choosing any specific form
of it. They found that $\Delta$ crosses the non-interacting line
($\Delta=0$) and changes its sign around $z=0.5$. This behavior of
interacting term raises remarkable challenge to the interacting
models since the above three forms of interaction do not changes
their signs during the cosmological evolution. This gives a clue to
the proposal of most general form of interaction.

This is why, Sun and Yue (2012) to proposed new form of interaction
term as follows
\begin{equation}\label{5}
\Delta_4=3\xi^2H(\rho_{\Lambda}-\rho_{m}),
\end{equation}
Here $\xi^2>0$ because its negative value conducts the matter energy
density to be negative. In the early universe,
$\rho_m>\rho_{\Lambda}$ leads to $\Delta_4<0$ while in present
epoch, it changes its sign after the transition of universe from
decelerated to accelerated regime. As, our universe shows transition
from deceleration to acceleration at the redshift value $z=0.5$. It
is also found that this interaction term shows the compatibly with
the universe transition at $z=0.5$. Also, the scenario corresponding
to the above interaction term satisfied the generalized second law
of thermodynamics. They have also compared this interacting model
with observational data and found its consistencies with the result
in (Daly et al. 2008) and the $7$-years WMAP observations (Komatsu
et al. 2011).

Differentiating Eq.(\ref{3}) with respect to $x=\ln a$, we obtain
\begin{equation}\label{6}
\rho'_{\Lambda}=u \rho_{\Lambda}\left(\frac{\alpha+2\beta
H}{\alpha+\beta H}\right)\frac{\dot{H}}{H^2}.
\end{equation}
where prime means differentiation with respect to $x$. Using
Eq.(\ref{4})-(\ref{6}), we obtain the EoS parameter as
\begin{equation}\label{7}
\omega_{\Lambda}=-1-\xi^2\left(1-\frac{\Omega_m}{\Omega_\Lambda}\right)-\frac{u}{3}\left(\frac{\alpha+2\beta
H}{\alpha+\beta H}\right)\frac{\dot{H}}{H^2}.
\end{equation}
In order to eliminate the term $\frac{\dot{H}}{H^2}$ from this
equation, we use Eqs.(\ref{1})-(\ref{6})
\begin{equation}\label{7a}
\frac{\dot{H}}{H^2}=\frac{(3\xi^2\Omega_{\Lambda}-\Omega_{m}(3\xi^2+1))(\alpha+\beta
H)}{2(\alpha+\beta H)-u\Omega_{\Lambda}(\alpha+2\beta H)}.
\end{equation}
Thus the EoS parameter becomes
\begin{equation}\label{7b}
\omega_{\Lambda}=-1-\xi^2\left(1-\frac{\Omega_m}{\Omega_\Lambda}\right)
-\frac{u(\alpha+2\beta)(3\xi^2\Omega_{\Lambda}-\Omega_{m}(3\xi^2+1))}{3(2(\alpha+\beta
H)-u\Omega_{\Lambda}(\alpha+2\beta H))}.
\end{equation}

We plot this EoS parameter versus PDE parameter $u$ for its three
different ranges $0< u\leq1.3$, $1.4\leq u\leq12$ and $-16\leq
u\leq0$ as shown in Figures \textbf{1-3}. To analyze the behavior of
EoS parameter more clearly, we choose three different well-known
values of interacting parameter $\xi^2=0,~0.5,~1$ and keep the
present values of other parameters such as
$\Omega_{\Lambda}=0.76,~\Omega_{m}=0.24$ (Suyu et al. 2013). Also,
the values of the generalized ghost version of PDE parameters are
$\alpha=1.55,~\beta=1.91$. In Figure \textbf{1} (for $\xi^2=0$), the
EoS parameter meets $\Lambda$CDM model for $0< u\leq0.4$, maintains
the quintessence DE region for $0.4< u\leq1.2$ and then goes towards
DM region of the universe. Consequently, this model does not favor
the phenomenon of PDE. For $\xi^2=0.5,~1$, the EoS parameter starts
from $-1.25,-1.6$ and goes towards more negative values. Thus EoS
parameter indicates the presence of less phantom energy at low
values of $u$ and converges towards strong phantom region with the
increase of $u$. This behavior strongly favors the PDE phenomenon
and also there is a possibility of big-rip singularity.
\begin{figure} \centering
\epsfig{file=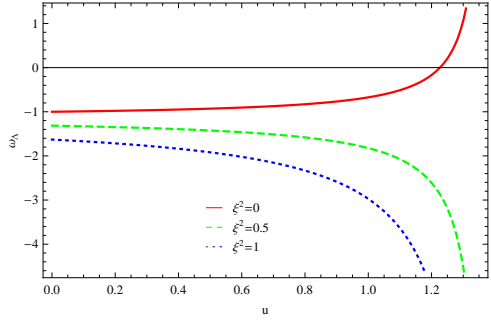,width=.50\linewidth}\caption{Plot of
$\omega_{\Lambda}$ versus $u$ for generlized ghost version of PDE
with $0\leq u\leq1.3$.}
\end{figure}
\begin{figure} \centering
\epsfig{file=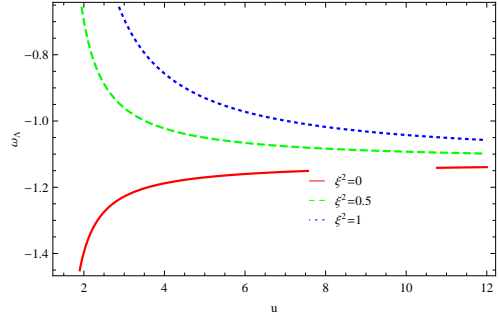,width=.50\linewidth}\caption{$\omega_{\Lambda}$
versus $u$ for generlized ghost version of PDE with $u\geq1.4$.}
\end{figure}
\begin{figure} \centering
\epsfig{file=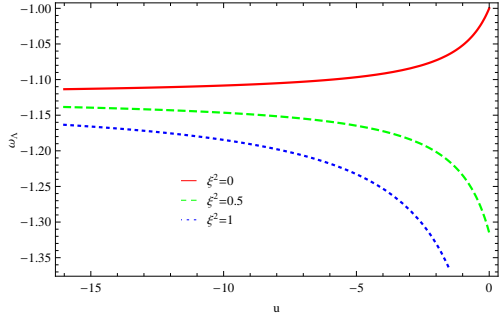,width=.50\linewidth}\caption{$\omega_{\Lambda}$
versus $u$ with $u\leq0$.}
\end{figure}

In Figure \textbf{2}, it is observed that the present values of EoS
parameter start from the high phantom region and approach to less
phantom region, i.e., $\omega_\Lambda=-1.15$ for $\xi^2=0$. When
$\xi^2=0.5,~\omega_\Lambda$ represents quintessence region in the
range $1.8\leq u<3.25$, represents the $\Lambda$CDM model at
$u=3.25$ and approaches to phantom era value $-1.1$ for $u>3.25$. In
case of $\xi^2=1$, the present values of $\omega_\Lambda$ start from
quintessence region and approach to phantom region by crossing the
$\Lambda$CDM limit. In this graph, all three models correspond to
phantom region which is useful for PDE prediction. In Figure
\textbf{3}, the model with $\xi^2=0$ coincides with $\Lambda$CDM
limit at $u=0$ and goes towards phantom region for $u<0$. For
$\xi^2=0.5,~1$, the present values of EoS parameter lie in the
phantom region for $u\leq0$.

Now, we use squared speed of sound for the stability analysis of the
present interacting model given by
\begin{equation}\label{15}
\upsilon_{s}^2=\frac{\dot{p}}{\dot{\rho}}=\frac{p'}{\rho'},
\end{equation}
Using Eqs.(\ref{3}), (\ref{4}), (\ref{6}), (\ref{7a}) and
(\ref{15}), it follows that
\begin{eqnarray}\nonumber
\upsilon^2_{s}&=&\frac{1}{3}(-3+3\xi^2(-2+\Omega_\Lambda^{-1})+(u(-1-3\xi^2+
\Omega_\Lambda+6\xi^2\Omega_\Lambda)(\alpha\\\nonumber &+&2\beta
H))(-2(\alpha+\beta H)+u \Omega_\Lambda(\alpha+2\beta
H))^{-1}+((\alpha+\beta H)(2(\alpha\\\nonumber &+&\beta
H)-u\Omega_\Lambda H(\alpha+2\beta H))(((-1+u)(12\xi^2(\alpha +\beta
H)^2-12\xi^2u\Omega_\Lambda\\\nonumber &\times&(\alpha+\beta
H)(\alpha+2\beta H)+u\Omega_\Lambda^2(\alpha+2\beta
H)(\alpha(2+12\xi^2-u)+2\beta(1\\\nonumber
&+&6\xi^2-u)H)))(-2(\alpha+\beta H)+u\Omega_\Lambda(\alpha+2\beta
H))^{-1}+(2\alpha\beta u \Omega_\Lambda\\\nonumber &\times&
(-1-3\xi^2+\Omega_\Lambda+6\xi^2\Omega_\Lambda)H)(-2\alpha+H(-2\beta
+u\Omega_\Lambda(\alpha+2\beta H)))^{-1}))\\\nonumber
&\times&(u\Omega_\Lambda(\alpha+2\beta H)(-2(\alpha+\beta
H)+u\Omega_\Lambda (\alpha+2\beta H))^2)^{-1}).
\end{eqnarray}
The behavior of $\upsilon_{s}^2$ against PDE parameter $u$ for its
three ranges (keeping other cosmological parameters the same) is
shown in Figures \textbf{4-6}. In Figure \textbf{4}, the
non-interacting generalized ghost version of PDE remains stable in
the ranges $0\leq u\leq0.25$ and $0.9\leq u\leq1.1$ while exhibits
instability in the ranges $0.25< u<0.9$ and $u>1.1$. However, the
models corresponding to interacting cases $\xi^2=0.5,~1$ show
stability in the ranges $0\leq u\leq0.02,~u\geq0.95$ and instability
for $0.02<u<0.95$. The squared speed of sound remains positive for
non-interacting case as shown in Figure \textbf{5}. However, its
behavior is similar for interacting cases, i.e, it starts from
negative values (represents instability of the models), goes towards
positive maxima and eventually decreases and approaches to positive
value (exhibits stability of the models). It is remarked that theses
models remain stable forever for $u\geq2.5,~u>1.6,~u>1.8$
corresponding to $\xi^2=0,~0.5,~1$ cases, respectively. Also, it is
observed that $v_s^2<0$ for $-0.1<u<0,~-0.8<u<0,~-1.5<u<0$ for
$\xi^2=0,~0.5,~1$, respectively which shows that these models
exhibit instability for these ranges of $u$ as shown in Figure
\textbf{6}. However, these models corresponding to $\xi^2=0,~0.5,~1$
remain stable for $u\leq-0.1,-0.8,-1.5$, respectively.
\begin{figure} \centering
\epsfig{file=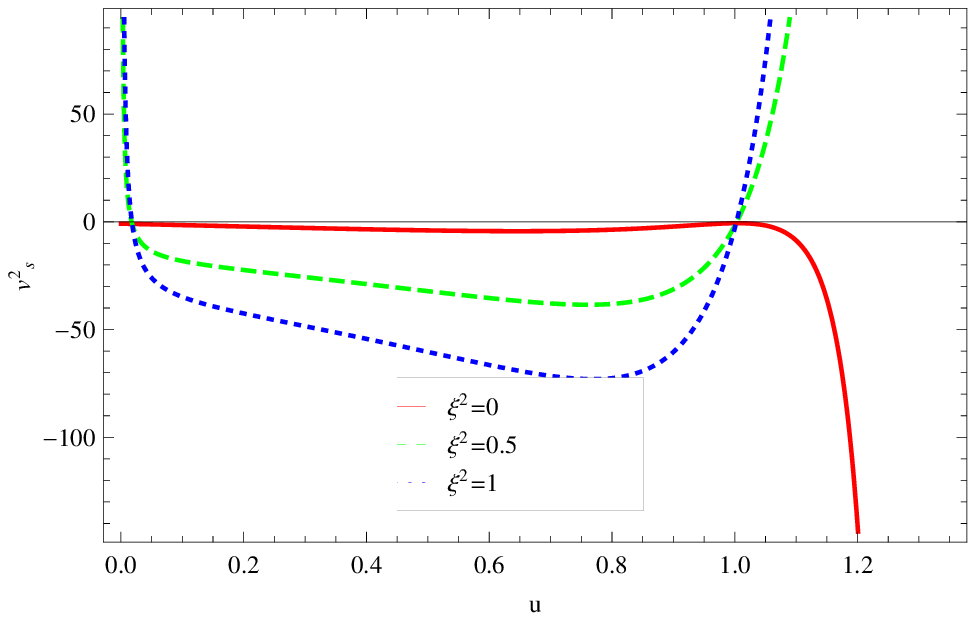,width=.50\linewidth}\caption{$v_s^2$ versus $u$
for generlized ghost version of PDE with $0\leq u\leq1.3$.}
\end{figure}
\begin{figure} \centering
\epsfig{file=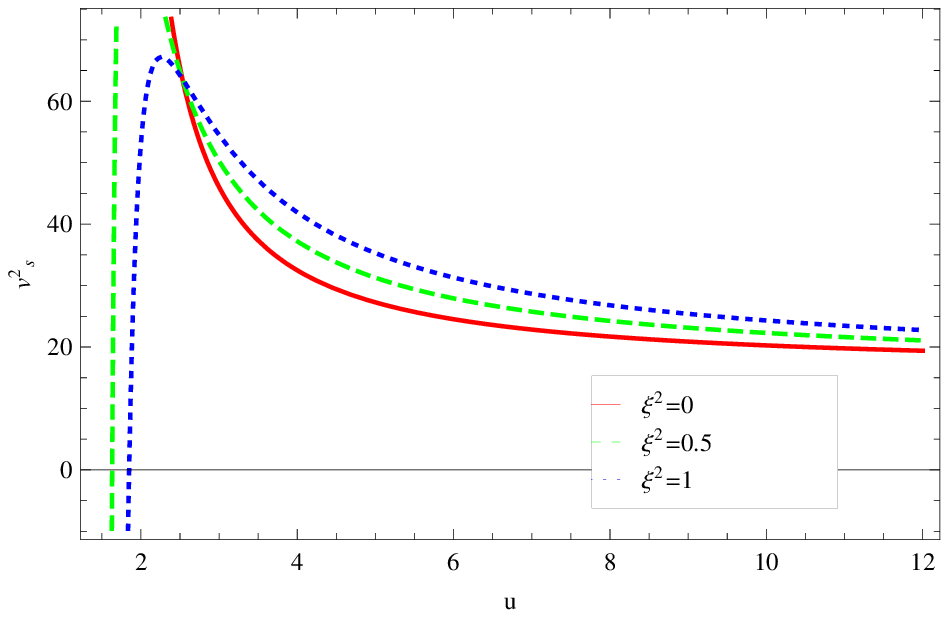,width=.50\linewidth}\caption{$v_s^2$ versus $u$
for generlized ghost version of PDE with $u\geq1.4$.}
\end{figure}
\begin{figure} \centering
\epsfig{file=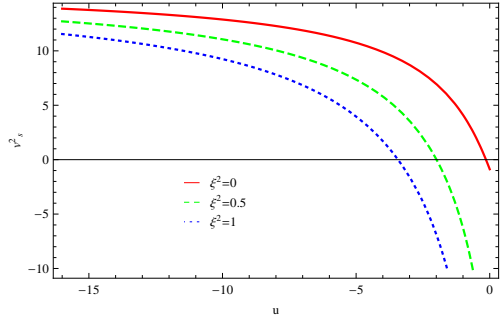,width=.50\linewidth}\caption{$v_s^2$ versus $u$
with $u\leq0$.}
\end{figure}

\section{$\omega_{\Lambda}-\omega'_{\Lambda}$ Analysis}

The $\omega_{\Lambda}-\omega'_{\Lambda}$ plane, firstly proposed by
Caldwell and Linder (2005), has become useful tool for
distinguishing different DE models through trajectories on its
plane. Initially, this approach has been applied on quintessence DE
model which leads to two classes of its plane, i.e, the area
occupied by the region ($\omega'_{\Lambda}>0,~\omega_{\Lambda}<0$)
on $\omega_{\Lambda}-\omega'_{\Lambda}$ plane corresponds to thawing
region while area under the region ($\omega'_{\Lambda}<0$,
$\omega_{\Lambda}<0$) implies the freezing region. It is observed
that the expansion of the universe is comparatively more
accelerating in freezing region. Later, this tool has been applied
to other well-known dynamical DE models such as more general form of
quintessence (Scherrer 2006), phantom (Chiba 2006), quintom (Guo et
al. 2006), polytropic DE (Malekjani and Khodam-Mohammadi 2012) and
PDE ( Sharif and Jawad 2013a, 2013b) (in flat and non-flat
universes) models. Here we use this analysis to explore these
regions. Differentiating $\Omega_{\Lambda}$ with respect to $x$ and
using Eqs.(\ref{1})-(\ref{3}) and (\ref{7a}), we obtain
\begin{eqnarray}\nonumber
\Omega'_\Lambda&=&\Omega_\Lambda[u(\alpha+2\beta H)-\alpha-\beta
H][-\Omega_m+3\xi^2(\Omega_\Lambda-\Omega_m)][2(\alpha\\\label{W1}
&+&2\beta H)-u\Omega_\Lambda(\alpha+2\beta H)]^{-1},
\end{eqnarray}
By taking the derivative of Eq.(\ref{7b}) and using the above
expression, we get the evolutionary form of $\omega_{\Lambda}$ as
follows
\begin{eqnarray}\nonumber
\omega'_{\Lambda}&=&(-((2\alpha\beta
u\Omega_{\Lambda}^2(-1-3\xi^2+\Omega_{\Lambda}+6\xi^2\Omega_{\Lambda})(-1
+\Omega_{\Lambda}+3\xi^2(-1\\\nonumber
&+&2\Omega_{\Lambda}))H(\alpha+\beta H))(2(\alpha+\beta H)-u
\Omega_{\Lambda}H(\alpha+2\beta
H))^{-1})-(\Omega_{\Lambda}\\\nonumber
&\times&(-1+\Omega_{\Lambda}+3\xi^2(-1+2\Omega_{\Lambda}))(-\alpha-\beta
H+u(\alpha+\beta H))(12\xi^2(\alpha
\\\nonumber
&+&\beta H)^2-12\xi^2 u\Omega_{\Lambda}(\alpha+\beta
H)(\alpha+2\beta H)+u \Omega_{\Lambda}^2(\alpha+2\beta H)\\\nonumber
&\times&(\alpha(2+12\xi^2-u)+2\beta(1+6\xi^2-u)H)))(2(\alpha+\beta
H)-u \Omega_{\Lambda}\\\label{W2} &\times&(\alpha+2\beta
H))^{-1})(3\Omega_{\Lambda}^2(-2(\alpha+\beta H)+u
\Omega_{\Lambda}(\alpha+2\beta H))^2)^{-1}.
\end{eqnarray}
\begin{figure} \centering
\epsfig{file=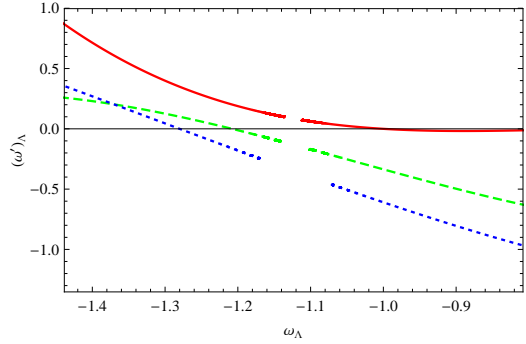,width=.50\linewidth}\caption{Plot of
$\omega_{\Lambda}-\omega'_{\Lambda}$ for generlized ghost version of
PDE. Red, green and blue lines correspond to $\xi^2=0,~0.5,~1$,
respectively.}
\end{figure}

We can obtain the $\omega_{\Lambda}-\omega'_{\Lambda}$ plane by
plotting $\omega'_{\Lambda}$ versus $\omega_{\Lambda}$ and keeping
the same values of constant cosmological parameters as shown in
Figure \textbf{7}. It can be observed that $\Lambda$CDM limit
($\omega_\Lambda'=0$ when $\omega_\Lambda=-1$) can be achieved for
non-interacting case. Also, the curve corresponds to the crossing
line of freezing and thawing regions, i.e., $\omega'_{\Lambda}=0$
upto the range $-1.02\leq\omega_\Lambda\leq-0.8$ while it goes
towards thawing region for $\omega_\Lambda\leq-1.02$. For
interacting case $\xi^2=0.5$, the curve does not meet the
$\Lambda$CDM limit. However, it remains in the freezing region in
the range $-1.2<\omega_\Lambda\leq-0.8$ and in the thawing region
for $\omega_\Lambda<-1.2$. When $\omega_\Lambda=-1.2,-1$,
$\omega'_{\Lambda}$ approaches to $0$ and $-0.3$, respectively. For
interacting case $\xi^2=1$, the freezing and thawing regions are
also obtained in the ranges $\omega_\Lambda<-1.28$ and
$-1.28<\omega_\Lambda\leq-0.8$, respectively while for
$\omega_\Lambda=-1.28,-1$, $\omega'_{\Lambda}$ approaches to $0$ and
$-0.6$, respectively.

\section{Statefinder Parameters}

Since now, a variety of DE models have been proposed for explaining
the accelerated expansion phenomenon of the universe. In order to
check the viability of these models, statefinder parameters are
widely used (Sharif and Jawad 2013a, 2013b; Sahni et al. 2003; Alam
et al. 2003; Feng 2008; Setare et al. 2007; Malekjani et al. 2011).
The cosmological plane corresponding to these parameters termed as
$r-s$ plane and the trajectories tell the distance of a given DE
model from $\Lambda$CDM limit. The statefinder parameters for flat
universe are defined as follows
\begin{eqnarray}\label{r1}
r=\frac{\dddot{a}}{aH^3},\quad s=\frac{r-1}{3(q-\frac{1}{2})},
\end{eqnarray}
where $q$ indicates the deceleration parameter. The cosmological
plane of these parameters describe different well-known regions of
the universe, i.e., $s>0$ and $r<1$ describe the region of phantom
and quintessence DE eras, $(r,s)=(1,0)$ corresponds to $\Lambda$CDM
limit, $(r,s)=(1,1)$ represents CDM limit and $s<0$ and $r>1$
indicates chaplygin gas. Using Eqs.(\ref{7a}) and (\ref{r1}), it
follows that
\begin{eqnarray}\nonumber
r&=&1-((-1+\Omega_{\Lambda}+\xi^2(-3+6\Omega_{\Lambda}))(\alpha+2\beta
H)u(-6(\Omega_{\Lambda}+\xi^2(-1\\\nonumber
&+&2\Omega_{\Lambda}))(\alpha+\beta
H)+\Omega_{\Lambda}(1+2\Omega_{\Lambda})(\alpha+2\beta
H)u))(2(-2(\alpha+\beta H)
\\\nonumber
&+&\Omega_{\Lambda}(\alpha+2\beta
H)u)^2)^{-1}-((-1+\Omega_{\Lambda}+\xi^2(-3+6\Omega_{\Lambda}))
(\alpha+\beta H)\\\nonumber
&\times&((2\alpha\Omega_{\Lambda}(-1+\Omega_{\Lambda}+\xi^2(-3+6\Omega_{\Lambda}))
\beta H u)(-2(\alpha+\beta H)+\Omega_{\Lambda}H\\\nonumber
&\times&(\alpha+2\beta H)u)^{-1}+((-1+u)(12\xi^2(\alpha+\beta
H)^2+2\Omega_{\Lambda}(6\xi^2(-1\\\nonumber
&+&\Omega_{\Lambda})+\Omega_{\Lambda})(\alpha+\beta H)(\alpha+2\beta
H)u-\Omega_{\Lambda}^2(\alpha+2\beta H)^2u^2))(-2(\alpha\\\nonumber
&+&\beta H)+\Omega_{\Lambda}(\alpha +2\beta
H)u)^{-1}))(2(-2(\alpha+\beta H)+\Omega_{\Lambda}(\alpha+2\beta
H)u)^2)^{-1},\\\label{r2}
s&=&((-1+\Omega_{\Lambda}+\xi^2(-3+6\Omega_{\Lambda}))(9(\alpha+2\beta
H)u-((\alpha+\beta H)(6(\alpha\\\nonumber &+& \beta
H)-3\Omega_{\Lambda} (\alpha+2\beta
H)u)^2((2\alpha\Omega_{\Lambda}(1+\Omega_{\Lambda}+\xi^2(-3+6\Omega_{\Lambda}))\\\nonumber
&\times&\beta H u)(-2(\alpha+\beta H)+\Omega_{\Lambda}
H(\alpha+2\beta H)u)^{-1}+((-1+u)(12\xi^2\\\nonumber
&\times&(\alpha+\beta H)^2+2\Omega_{\Lambda}(6\xi^2(-1
+\Omega_{\Lambda})+\Omega_{\Lambda})(\alpha+\beta H)(\alpha+2\beta
H)u\\\nonumber &-&\Omega_{\Lambda}^2(\alpha+2\beta
H)^2u^2))(-2(\alpha+\beta H)+\Omega_{\Lambda}(\alpha+2\beta
H)u)^{-1}))((-2(\alpha\\\nonumber &+&\beta
H)+\Omega_{\Lambda}(\alpha+2\beta
H)u)^2(6(\Omega_{\Lambda}+\xi^2(-1+2\Omega_{\Lambda}))(\alpha+\beta
H)\\\nonumber &-&\Omega_{\Lambda}(1+2\Omega_{\Lambda})(\alpha+2\beta
H)u))^{-1}))(9(-6(\alpha+\beta
H)+3\Omega_{\Lambda}(\alpha\\\label{r3} &+&2\beta H)u))^{-1}.
\end{eqnarray}
\begin{figure} \centering
\epsfig{file=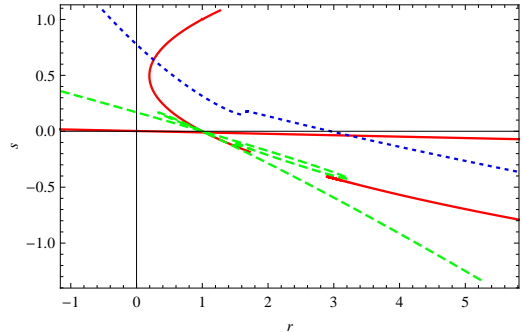,width=.50\linewidth}\caption{Plot of $r-s$ for
generlized ghost version of PDE. Red, green and blue lines
correspond to $\xi^2=0,~0.5,~1$, respectively.}
\end{figure}

We can obtain the plane of statefinders by plotting $s$ versus $r$
for three different choices of interacting parameter as shown in
Figure \textbf{8}. It can be observed from the $r-s$ plane that
non-interacting and interacting ($\xi^2=0.5$) generalized ghost PDE
models correspond to $\Lambda$CDM model. The $r-s$ plane
corresponding to three different choices of interacting parameter
($\xi^2=0,~0.5,~1$) also provide the regions of DE (phantom and
quintessence) and chaplygin gas model. In non-interacting case, the
trajectory also corresponds to CDM limit.

\section{Concluding Remarks}

The generalized ghost DE model has been used for different purposes
such as the evolution of the universe by extracting different
cosmological parameters, thermodynamics laws, correspondence with
different scalar field models, analysis of
$\omega_\Lambda-\omega'_\Lambda$ and $r-s$ planes. Also, various
cosmological aspects of this model have been addressed in different
modified theories. This idea of PDE is interesting as it indicates
one of the notions about the universe (phantom energy) in the later
time. In this work, we have developed generalized ghost version of
PDE to explain the fate of BH in the presence of large amount of
phantom energy in the universe.

We have considered interacting scenario of the generalized ghost DE
with CDM and evaluated two cosmological parameters (i.e., EoS and
squared speed of sound) as well as two cosmological planes (i.e.,
$\omega_\Lambda-\omega'_\Lambda$ and $r-s$). We have analyzed the
behavior of these parameters through two constant parameters such as
interacting ($\xi^2$) and PDE ($u$). We have explored EoS parameter
versus PDE parameter for three different well-known values of
$\xi^2=0,~0.5,~1$. Also, we have taken into account three ranges of
PDE parameter, i.e., $0\leq u\leq1.3,~u\geq1.4$ and $u\leq0$ as
shown in Figure \textbf{1-3}. It can be observed from Figure
\textbf{1} that the present values of $\omega_\Lambda$ do not lie in
the phantom region of the universe for non-interacting case and
hence it does not favor the PDE conjecture. In the interacting case,
the present values of $\omega_\Lambda$ correspond to the phantom
region and attain more negative (phantom) values with the increase
of $u$. However, the presence of phantom energy for the interacting
case ($\xi^2=1$) is comparatively larger than the interacting case
($\xi^2=0.5$).

Figure \textbf{2} shows that all the present values of EoS parameter
in the non-interacting case lie in the phantom region and it attain
the region of high phantom energy at $u=1.4$. Moreover, the present
values of $\omega_\Lambda$ approach to phantom region for $u\geq3$
and $u\geq8$ in the interacting cases $\xi^2=0.5~1$, respectively.
For $u<0$ (Figure \textbf{3}), the trajectory of $\omega_\Lambda$
corresponding to all the cases of interacting parameter lies in the
phantom region. However, the interacting model with $\xi^2=1$
acquires more phantom energy as compared to other cases. For the
stability analysis of these models, we have plotted the squared
speed of sound versus $u$ (with three different ranges) for the
non-interacting and interacting cases as shown in Figures
\textbf{4-6}. It is found that the generalized ghost PDE exhibits
stability in the ranges $u\leq-0.1,~0\leq u\leq0.25,~0.9\leq
u\leq1.1$ and $u\geq2.5$ for non-interacting case. For interacting
case ($\xi^2=0.5$), this PDE model remains stable in the ranges
$0\leq u\leq0.02,~u\geq1.6$ and $u\leq-0.8$. For $\xi^2=1$,
stability of the model is observed when $u>1.1,~u>1.8$ and
$u\leq-1.5$.

We have also examined the consequences of
$\omega_{\Lambda}-\omega'_{\Lambda}$ plane for this PDE model which
shows that the $\Lambda$CDM limit is achieved only for
non-interacting case. We have also mentioned different ranges of PDE
parameter $u$, where the $\omega_{\Lambda}-\omega'_{\Lambda}$ plane
corresponds to thawing and freezing regions for non-interacting as
well as interacting case. Finally, we have developed $r-s$ plane and
found that the trajectories corresponding to non-interacting and
interacting ($\xi^2=0.5$) cases meet the $\Lambda$CDM limit. It is
also pointed out that the $r-s$ plane for non-interacting and
interacting cases possess the regions of chaplygin gas, quintessence
and phantom models. We conclude that this work favors the PDE
phenomenon.

\end{document}